\documentclass[superscriptaddress,amsmath,amssymb,aps,prb,twocolumn]{revtex4-1}
\usepackage{dcolumn}
\usepackage{hyperref} 
\usepackage{graphicx}
\usepackage{amsmath}
\usepackage{siunitx}
\usepackage{xcolor}
\usepackage{bm}
\usepackage{textcomp} 
\usepackage{chemformula}

\begin{document}

\title{Superconducting phase transition in inhomogeneous chains of superconducting islands}

\author{Eduard Ilin}
\affiliation{Department of Physics, University of Illinois at Urbana-Champaign, Urbana, IL 61801, USA}
\author{Irina Burkova}
\affiliation{Department of Physics, University of Illinois at Urbana-Champaign, Urbana, IL 61801, USA}
\author{Xiangyu Song}
\affiliation{Department of Physics, University of Illinois at Urbana-Champaign, Urbana, IL 61801, USA}
\author{Michael Pak}
\affiliation{Department of Physics, Air Force Institute of Technology, Wright-Patterson AFB, Dayton, OH 45433, USA}
\author{Dmitri S. Golubev}
\affiliation{Pico group, QTF Centre of Excellence, Department of Applied Physics, Aalto University, FI-00076 Aalto, Finland}
\author{Alexey Bezryadin}
\affiliation{Department of Physics, University of Illinois at Urbana-Champaign, Urbana, IL 61801, USA}

\date{\today}

\begin{abstract}
We study one dimensional chains of superconducting islands with a particular emphasis on the regime in which every second island is switched into its normal state, thus forming a superconductor-insulator-normal metal (S-I-N) repetition pattern. As is known since Giaever tunneling
experiments \cite{Giaever}, tunneling charge transport between a superconductor
and a normal metal becomes exponentially suppressed, and zero-bias resistance diverges, as the temperature
is reduced and the energy gap of the superconductor grows larger than the thermal energy. Here we demonstrate
that this physical phenomenon strongly impacts transport properties of inhomogeneous superconductors made of
weakly coupled islands with fluctuating values of the critical temperature. We observe a non-monotonous
dependence of the chain resistance on both temperature
and magnetic field, with a pronounced resistance peak at temperatures at which some but not all islands are superconducting.
We explain this phenomenon by the inhomogeneity of the chains, in which neighboring superconducting
islands have slightly different critical temperatures. We argue that the Giaever's resistance divergence
can also occur in the zero-temperature
limit. Such quantum transition can occur if the magnetic field is tuned such that it suppresses superconductivity in the islands
with the weaker critical field, while the islands with stronger energy gap remain superconducting. In such a field, the system acts as a chain of S-I-N junctions.
\end{abstract}

\maketitle

\section{Introduction}
Typically, when a metallic film or a wire is cooled below the critical temperature of the superconducting phase transition, its resistance monotonically
decreases and approaches zero at sufficiently low temperatures. The width of the phase transition is usually determined by fluctuations of
the superconducting order parameter, spontaneous creation of vortices or phase slips, or by inhomogeneity of the sample.
Surprisingly, in some cases, especially in mesoscopic samples with a strong disorder, zero bias resistance grows above the normal resistance before it drops into the superconducting regime. Thus, a pronounced peak in the temperature dependence of the resistance is formed\cite{Dynes,Hebard,Hebard2,Adkins,Briceno,Yurgens,Gantmakher}.

We study a model system showing this puzzling behavior and explain the phenomenon by taking into account the chain inhomogeneity leading to an alteration of superconducting and normal elements. The phenomenon is similar in nature to the resistive peak reported by Giaever \cite{Giaever} in superconductor-normal metal tunnel junctions and related to the mismatch of the electronic energy gap values in such junctions. The generality and the importance of this phenomenon is based on the fact that any granular thin film, made of small grains separated by oxidized layers or some other tunnel barriers should show a similar behavior. We show that if S-I-N type patterns are present then the resistance is expected to diverge \cite{ChaudhuriBagwell}. Such weak coupling between normal and superconducting lakes can also occur in films or thin wires composed of a mixture of metallic atoms and oxygen atoms, where random potential fluctuations are strong \cite{Gantmakher}. Here we propose a unifying view to explain the origin of the ubiquitous resistance peak in the resistance versus temperature and/or resistance versus magnetic field dependences.

As a side note, we speculate that the physics of the Giaever resistance divergence, might even be applicable to strongly disordered but homogeneous amorphous superconducting films. The basis of this hypothesis is a strong inhomogeneity of the superconducting energy gap, previously discovered in pioneering experiments of Sacépé et al. \cite{SacepeDisorder,SacepeSI}. In these experiments it was also discovered that a substantial fraction of the sample can be in the insulating state, where the energy gap does exist, but the divergence of the density of states at the gap edges, expected in superconductors, cannot be observed. If a sufficient percentage of the sample is insulating, then the tunneling barriers can emerge between the superconducting and/or normal regions. An S-I-N pattern can develop if the temperature and/or the magnetic field are adjusted such that some but not all superconducting segments of the samples are converted into the normal states. A diverging resistance is expected in such regime.

Here we report the experimental study, accompanied by the theory analysis, of the resistance peak effect in very long S-I-S' chains. Previously, such chains have been investigated at microwave frequency \cite{Kuzmin}, and here we focus on the dc transport.
The chain consists of aluminum islands connected by tunnel junctions with relatively high resistance of the order of the quantum resistance. In our experiment, we observed an increase of the sample resistance above the normal state value at a certain temperature and/or magnetic field intervals.
We explain the observed non-monotonous temperature dependence of the zero-bias resistance by the difference in the critical temperature of the even and the odd
superconducting islands. By lowering the temperature, let us say, odd-number islands switch to the superconducting state while the neighboring, even-number islands, still remain in the normal state.
As a result, in the temperature interval between the two critical temperatures, one obtains a one-dimensional chain of normal metal-insulator - superconductor (S-I-N) Giaever junctions.
Zero bias resistance of such junctions is known to always grow with lowering temperature due to the opening of the gap in the density of states
of the superconductor \cite{Giaever}. As soon as all the islands convert to the superconducting state, we obtain a chain of Josephson S-I-S' junctions, and its zero-bias resistance
begins to decrease with cooling. We will demonstrate that this simple model well describes our observations of the resistance peak.

We also extrapolate our data towards zero temperature and argue that a superconductor-insulator quantum transition or a crossover can occur if different islands or regions of the sample have different critical magnetic field values. Such a case is quite possible in disordered mesoscopic samples exhibiting strong gap fluctuations\cite{SacepeDisorder,SacepeSI}. So, even at zero temperature, one can expect that, as the magnetic field is increased, the regions (islands) with the lower critical temperature will become normal, while the regions with a higher critical temperature will remain superconducting. The resistance of such an S-I-N chain should diverge. Thus, a zero temperature crossover or a quantum transition can be envisioned. We present indirect evidence of such transition in our chains of islands with alternating critical temperatures.

\section{Sample fabrication and measurements}
\label{sample}

\begin{figure}
\includegraphics[width=\columnwidth]{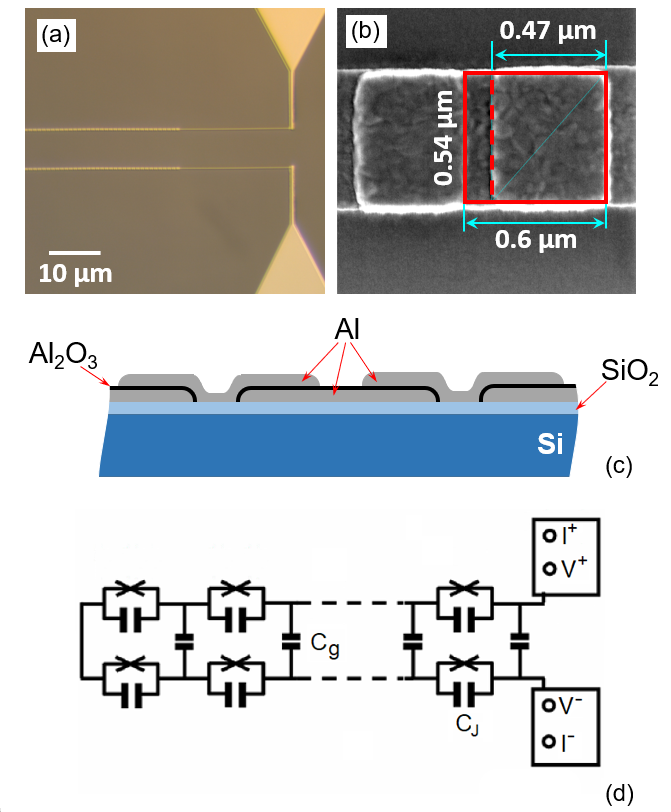}
\caption{(a) Optical image of the sample 053118SB. The contact pads are visible on the right side of the sample. The two parallel chains are connected at the bottom (not shown). (b) Scanning electron microscope (SEM) micrograph of the same sample. Individual Al islands are visible here. (c) Schematic side view (cross section) of the sample. The tunnel barrier between the islands is shown as a black curve. This tunnel barrier is made of the oxide, Al$_2$O$_3$, grown on the surface of the grains in a controlled oxidation procedure. This schematic illustrates the fact that the chain is made of the islands of  two types: The first type includes the bottom islands, i.e., those located directly on the oxidized Si wafer. The second type includes the Al islands deposited over the Al islands of the first type. Thus, the substrate for the second type islands is the oxidized aluminum and not the oxidized Si. (d) Electrical scheme of the sample. The sample contains 33000 SIS junctions organized into two parallel chains, mutually connected at the left end. The capacitance between the superconducting islands is $C_J$, and the mutual chain-to-chain capacitance is $C_0$, per island.}
\label{Fig1}
\end{figure}

We study long chains of Al islands coupled by tunnel junctions, made of $N=33000$ rectangular islands in total (Fig. \ref{Fig1}a), which is possibly the largest number tested so far. Such chains serve as a realistic model for a one-dimensional (1D) conductor with alternating local critical temperatures, $T_C$. The $T_C$ alternation is achieved since the Al film critical temperature is dependent on the substrate. In our chains, the islands make tunnel junctions with their neighbors by overlapping them. Thus, in each junction, one Al electrode rests on the substrate while the other one rests on the first Al electrode. Consequently, the critical temperatures of the banks of each tunnel junction differ by a few percent. We have done tests and explicitly confirmed this sensitivity of the critical temperature to the substrate: Al films deposited on sapphire and on another oxidized Al film showed different $T_C$ values.

The chains were fabricated using the standard Dolan bridge technique involving MMA/PMMA bi-layer resist patterned by electron beam lithography with subsequent double-angle deposition of aluminum with an intermediate oxidation step. Due to the large number of junctions in the chain, patterning was done by stitching multiple fields of view (100 $\mu$m each) of the electron-beam writing system. Two samples have been analyzed in detail 053118SB (Sample “B”) and 082408SE (Sample “E”). The width of the islands was 0.54 $\mu$m in Sample B (Fig. \ref{Fig1}b) and 0.3 $\mu$m in Sample E. The unit cell repetition length of each chain was 0.6 $\mu$m. The substrate for all samples was a high-resistivity silicon wafer. Since Si is a semiconductor and its energy gap is much larger than the thermal energy at the temperature of our experiment, the substrate behaves as an insulator.

\begin{figure}
\includegraphics[width=0.9\columnwidth]{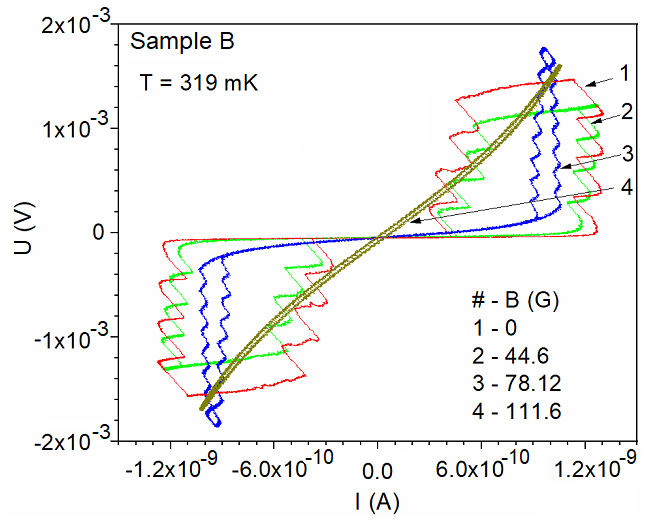}
\caption {Voltage-current (VI) curves of a the sample B. The parameter is the magnetic field. Here and everywhere the magnetic field was applied perpendicular to the substrate plane.}
\label{Fig2}
\end{figure}

Each sample was configured as a double-chain line (Fig. \ref{Fig1}a,\ref{Fig1}d), shorted at one of its ends, so that dc transport current can pass through the entire chain. We estimated the capacitance of each tunnel junction, $C_J$, by using specific capacitance\cite{David} of 45 fF/$\mu$m$^2$ and the junction area (0.47$\mu$m x 0.54$\mu$m) from scanning electron microscope (SEM) images (Fig. \ref{Fig1}b). The resulting values are $C_J= $11.4 fF and 6.7 fF for sample B and sample E, respectively.
Accordingly, the estimated values of the Coulomb charging energies are $E_C=e^2/2C_J$ = 7 and 12 $\mu$eV for samples B and E, respectively. The lowest temperature in the reported experiments was $T= $318 mK, which corresponds to the thermal energy 27 $\mu$eV. The inter-chain charging energy is 
defined as $E_{Cg} = e^2/2C_g$, where $C_g$ is the capacitance between each island and the other chain. The charging energy $E_{Cg}$, which is related to the island self capacitance, plays a key role in the ultimate long-scale and zero-temperature behavior in the theory \cite{Bard}. In particular, a JJ chain is expected \cite{BD,Bard,Lee,Kuzmin} 
to undergo a quantum superconductor-insulator transition at $K_{0}=\sqrt{E_{J}/8E_{Cg}}\sim 1$, where $E_{J}$ is per junction Josephson energy. We estimate that for sample B, $C_g=$22.82 aF, $E_{Cg}=$ 3.509 meV, $E_{J}=$ 365 $\mu$eV, $K_{0}=$ 0.114. For sample E, $C_{g}=$ 20.53 aF, $E_{Cg}=$ 3.898 meV, $E_{J}=$ 192 $\mu$eV, $K_{0}=$ 0.078. Thus, both arrays should, in principle, show insulating behavior at very low temperatures 
$k_BT\lesssim \hbar\omega_p\exp[-\sqrt{8E_J/E_C}]\sim 1$ mK, which we, however, do not explore here.
In this expression, $\omega_p$ is the plasma frequency of a junction defined as $\omega_p=\sqrt{8E_JE_C}/\hbar$. For samples B and E we estimate, correspondingly, $\hbar\omega_p=$144 $\mu$eV and $\hbar\omega_p=$135 $\mu$eV. The crossover temperature from thermal activation to tunneling between the wells of
the ``washboard'' Josephson potential of a single junction, $T^*=\hbar\omega_p/2\pi k_B$, takes the value 
$T^*=266$ mK  for the sample B and $T^*=249$ mK for the sample E, which is below
the lowest temperature in the reported experiments, 318 mK. Thus, in both samples thermally activated phase slips should give
the main contibution to zero bias resistance. 

DC transport measurements on the chains have been conducted in a $^3$He cryostat. The sample was installed in a Faraday cage. The resistive measurement leads, before entering the cage, were thermalized and filtered against external electromagnetic noise by cryogenic Cu powder and Ag powder filters. Additional $\pi$-filters have been installed on the cryostat room temperature input leads. The current bias of the sample was generated by a function generator DS360, which supplied a periodic voltage at a frequency of either 0.1 or 1 Hz. The voltage was applied to a standard resistor (1 M$\Omega$) connected in series with the sample. The voltages on the resistor and on the sample are amplified by dedicated PAR113 amplifiers and digitized. The voltage on the resistor was then converted to the current in the circuit using Ohm's law. The voltage-current (V-I) curves are plotted in the LabVIEW environment; the slope of the V-I curve measured near zero bias equals the sample zero-bias resistance. The magnetic field is generated by a superconducting solenoid and was always oriented perpendicular to the substrate surface, and, therefore, perpendicular to the planes of the overlap-type Josephson junctions. In other words, the magnetic field was always parallel to the tunneling supercurrent in the junctions. The temperature was measured and controlled using Lake Shore Cryotronics (LSC) 370AC system, connected to a commercially calibrated RuO thermometer, also supplied by LSC. The thermometer was placed near the sample into the same socket, such that its cooling was achieved through the leads, same way as the sample cooling.

Example V-I curves for sample B are shown in Fig. \ref{Fig2}. Similarly to the previous reports,\cite{David,David2}, the V-I curves show a series of almost identical steps. The steps are due to phase slip centers nucleating at individual junctions in the chain when the applied current exceeds the critical current of the junction. It is known that the size of the step is $2\Delta$, where $\Delta$ is the superconducting energy gap \cite{David}. The external magnetic field suppresses the critical current, making the V-I curve almost linear, as is illustrated in Fig. \ref{Fig2}.

\begin{figure*}
\includegraphics[width=1.7\columnwidth]{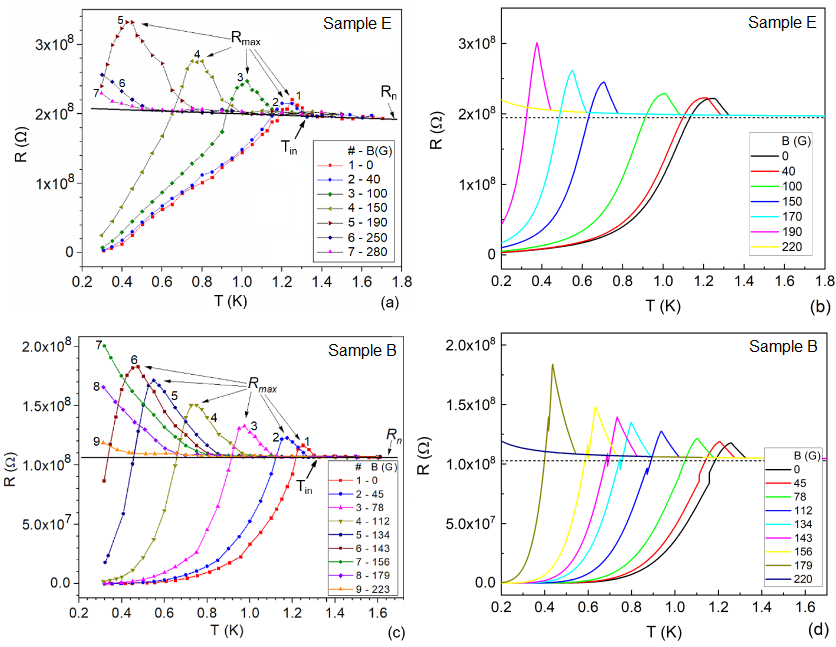}
\caption{Zero bias resistance versus temperature for the samples E (a,b) and B (c,d).
Left panels (a,c) show the measurements and the right panels - theory predictions, Eq. (\ref{R_par},\ref{Rqp},\ref{GJ},\ref{activation}).
The parameters used for the theory plots for sample E were
$E_C=e^2/2C_J=5.98$ $\mu$eV (which corresponds to the junction capacitance $C_J=13.4$ fF, two times more than the estimated capacitance $C_J=6.7$ fF), 
$B_{C1}=233$ G, $B_{C2}=227$ G,
$T_{C1}=1.325$ K, $T_{C2}=0.95 T_{C1}=1.259$ K, $N=33000$, the resistance
of a single junction is $R_n=5857$ $\Omega$, and the total normal state resistance of the array is $R_N=NR_n=1.933\times 10^8$ $\Omega$.
The theory parameters for sample B have been chosen as follows:
$E_C=e^2/2C_J=7.15$ $\mu$eV (which corresponds to the junction capacitance $C_J=11.2$ fF), $B_{C1}=233$ G, $B_{C2}=221$ G,
$T_{C1}=1.325$ K, $T_{C2}=0.95 T_{C1}=1.259$ K, $N=33000$, the resistance
of a single junction is $R_n=3118$ $\Omega$, and the total normal state resistance of the array is $R_N=NR_n=1.029\times 10^8$ $\Omega$.}
\label{RT}
\end{figure*}

The arrays exhibit a resistance peak as the temperature is reduced, as shown in Fig. \ref{RT}a, and Fig. \ref{RT}c. At the peak, the resistance is larger than the normal resistance of the sample. The peak effect is present even at zero magnetic field, yet it becomes larger, both in height and in width, if a perpendicular magnetic field is applied. We also find that at even higher magnetic fields, where the superconductivity is completely suppressed, the peak is not present, which indicates that the peak is related to the presence of the superconducting condensate in the system. A detailed model will be presented below.

As the temperature is further reduced, the resistance starts to go down quickly, as all islands enter their superconducting state. Yet, even at the lowest temperature tested, the samples exhibit some residual resistance. As will be analyzed in the next section, the sample resistance, in the temperature interval tested, can be explained by thermally activated phase slips occurring in the junctions of the chain.

\section{Zero bias resistance at zero magnetic field}

Here we present zero-bias resistance results measured at low temperatures 
and use the simplest possible model of thermally activated phase slips (TAPS) to analyze them. 
In this section we ignore Coulomb blockade effects, they will be discussed later. 
We assume that all islands are superconducting and that the junctions act as independent resistive elements (due to TAPS) connected in series.

The probability for a single junction in the chain to experience a phase slip is defined by the Arrhenius activation formula:
\begin{eqnarray}
\Gamma_{\rm jct}(I) = \frac{\omega_{p}(I)}{2 \pi } \exp \left[ -\frac{\Delta U(I)}{k_{B}T} \right].
\label{Gamma1}
\end{eqnarray}
Here $I$ is the bias current, $\omega_{p}(I)$ is bias dependent plasma frequency of the junction and $\Delta U(I)$ is the potential barrier
for a phase slip event. 
For each junction we use a tilted washboard potential $U(\varphi)=-(\hbar I_C/2e)\cos\varphi-(\hbar I/2e)\varphi$. 
The minimum of this potential occurs at $\varphi_{\min}=\arcsin(I/I_C)$ and the subsequent energy maximum --- at $\varphi_{\max}=\pi-\arcsin(I/I_C)$. 
Thus, the energy barrier for the phase slips in the positive direction is
\begin{eqnarray}
\Delta U_{+} &=& U(\varphi_{\max})-U(\varphi_{\min})
\nonumber\\
&=& \frac{\hbar}{e}\left[ \sqrt{I_C^2-I^2} + I\arcsin\frac{I}{I_C} - \frac{\pi I}{2}\right].
\end{eqnarray}
In the same way, one finds the energy barrier for the anti-phase slips, in which the phase revolves in the opposite direction,
\begin{eqnarray}
\Delta U_{-} = \frac{\hbar}{e}\left[ \sqrt{I_C^2-I^2} + I\arcsin\frac{I}{I_C} + \frac{\pi I}{2}\right].
\end{eqnarray}
The critical current for a single junction, $I_{C}$, is given by Ambegaokar-Baratoff formula \cite{AB}
\begin{eqnarray}
I_{C}=\frac{\pi}{2} \frac{\Delta(T)}{eR_{n}} \tanh \frac{\Delta(T)}{2k_{B}T},
\label{Ic}
\end{eqnarray}
where $R_{n}$ is the normal resistance of the junction. In this section, we use $R_{n}$ as well as the critical temperature as fitting parameters.
The dependence of the superconducting gaps of the islands follows Bardeen-Cooper-Schrieffer (BCS) theory \cite{BCS}, which we will approximate by an analytic expression:
\begin{eqnarray}
\Delta_j(T) = \pi e^{-\gamma} k_BT_{Cj}\tanh\left(\sqrt{\frac{8 e^{2\gamma}}{7\zeta(3)}}\sqrt{\frac{T_{Cj}}{T}-1}\right),
\label{DH}
\end{eqnarray}
where the subscript $j=1,2$ distinguishes the two types of the islands in the chain, 
$\zeta(x)$ is the Riemann zeta function and $\gamma=0.5772$ is the Euler–Mascheroni constant. Since this section is focused on the low-temperature part of the resistance versus temperature curve, we neglect small differences in the critical temperature and use one value, which represents the average critical temperature, i.e., we treat $T_{C1}=T_{C2}=T_{C}$ for the islands.
Finally, the plasma frequency of the junction, appearing in Eq. (\ref{Gamma1}), depends on the bias current as:
\begin{eqnarray}
\omega_{p}(I) = \frac{1}{\hbar} \sqrt{\frac{4E_{J}e^{2}}{C_{J}+C_{g}/4}} \left( 1 - \frac{I^{2}}{I_{C}^{2}} \right)^{1/4}.
\end{eqnarray}
Here $E_J=\hbar I_C/2e$, 
$C_{J}$ is the capacitance of each tunnel junction, and $C_{g}$ is the capacitance of the island to the ground or the the other, parallel chain.  
Since in both samples we find $C_g\ll C_{J}$, we neglect $C_g$ when computing the plasma frequency.

Having determined the rates for a single junction, we find
the rates for the phase slips and anti-phase slips in the entire array by simply multiplying them with the number of junctions,
\begin{eqnarray}
\Gamma_{\pm}(I)=N\frac{\omega_{p}(I)}{2\pi}\exp\left(-\frac{\Delta U^{\pm}(I)}{k_{B}T}\right).
\end{eqnarray}
Here $N=33000$ is the number of islands in the chain.

In the limit of low current bias the plasma frequency becomes independent of $I$,
\begin{eqnarray}
\omega_{p}(0) \approx \frac{2e}{\hbar}\sqrt{\frac{E_{J}}{C_{J}}} =\frac{1}{\hbar}\sqrt{\frac{e\Phi_{0}}{C_{J}}\frac{\Delta(T)}{R_{n}} \tanh \frac{\Delta(T)}{2k_{B}T}},
\end{eqnarray}
where $\Phi_0=\pi\hbar/e$ is the flux quantum.
In this regime, the voltage is expressed as 
\begin{eqnarray}
V(I)&=&\frac{\pi\hbar}{e}\left[\Gamma_{+}(I)-\Gamma_{-}(I) \right]
\nonumber\\
&=& N\frac{\hbar\omega_{p}(0)}{e}\exp\left(-\frac{2E_J}{k_{B}T}\right)\sinh\frac{\pi\hbar I}{2ek_BT},
\label{PSR1}
\end{eqnarray}
and the zero bias resistance takes the form
\begin{eqnarray}
R_{0}=\lim_{I\to 0}\frac{V(I)}{I} = NR_q\frac{\hbar\omega_{p}(0)}{4k_BT}\exp\left(-\frac{2E_J}{k_{B}T}\right). 
\label{R0}
\end{eqnarray}
Here $R_q=h/e^2$ is the resistance quantum. 

\begin{figure}
\includegraphics[width=0.9\columnwidth]{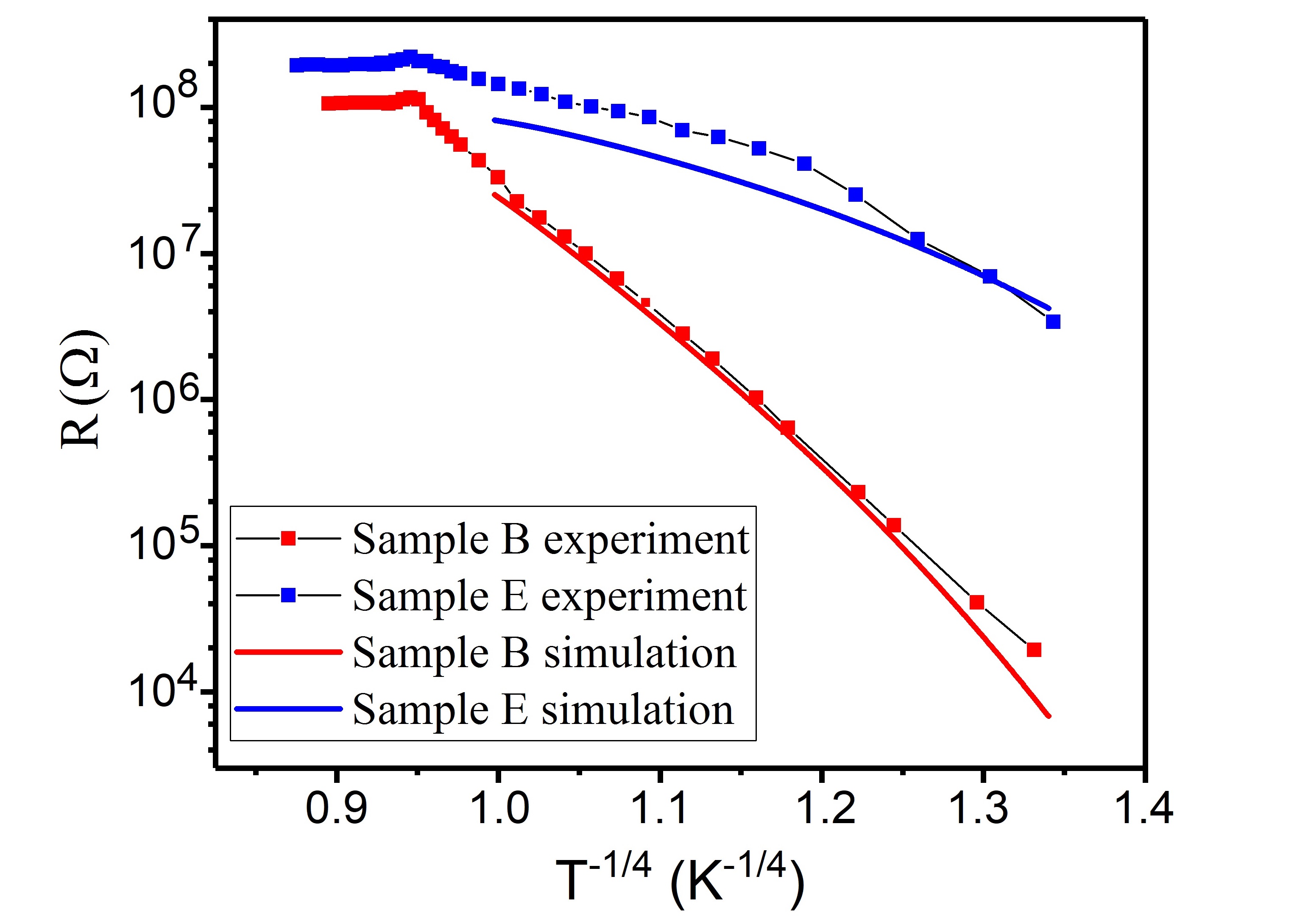}
\caption{Zero bias resistance versus temperature for the samples E and B. The experimental points are shown by the squares while the simple thermal phase slip model is shown by the continuous curves. The fitting parameters for sample B are: $R_{n}$=4.18 $k\Omega$, $T_{C}$=1.33$K$, and $C_{J}=$ 11.2 fF. For sample E the parameters are: $R_{n}$=9.435 $k\Omega$, $T_{C}$=1.39$K$, and $C_{J}=$ 3.5 fF.}
\label{fit}
\end{figure}

We use the expressions presented above in order to estimate the thermal phase slip rates in our arrays.
We find that these rates are high, even at the lowest temperature achieved in the experiment.  
Namely, the zero bias rate is found to be $10^{10}$ s$^{-1}$ for the sample B and $5\times 10^{12}$ s$^{-1}$ for the sample E. 
Simple numerical simulation showed that the net phase slip rate does not depend significantly on the bias current if it is less than 1 nA. 
Accordingly, in order to calculate zero bias resistance, we use a very small bias current of 0.1 nA.

In Fig. \ref{fit}, we present a comparison between the simple model presented in this section and the experimental results. 
The best fit paramters are indicated in the figure caption. The agreement is good for the sample B and also satisfactory for the sample E, 
taking into account the simplicity of the model. Yet, the fitting parameters are notably different from their expected values, 
which have been obtained independently. In the next section, we present a more advanced model, which gives better agreement with our experimental results.

\section{Superconducting transition in magnetic field}
\label{theory}

The resistance versus temperature, $R(T)$, curves measured at various magnetic fields are shown in Fig. \ref{RT}a, and Fig. \ref{RT}c. 
In this section, we present more advanced theoretical model, which allows us to fit $R(T)$ curves in a better way. The result of this model is plotted in Fig. \ref{RT}b, and Fig. \ref{RT}d,  and shows qualitatively similar behavior to the experimental curves. In particular, the model reproduces the key result --- the resistance peak. It also shows that the peak grows with increasing magnetic field.

In what follows, we present the model in some detail. First, we outline the main results related to superconducting tunnel junctions. We will neglect possible self-capacitance of the islands in the chain, and treat the tunnel junctions as independent. We assume for now that the magnetic field is absent. The key property of the considered chain, causing the resistance peak, is that neighboring superconducting islands have slightly different critical temperatures, namely,
the islands with odd numbers have the critical temperature $T_{C1}$ and the islands with even numbers --- $T_{C2}$,
and $T_{C1}>T_{C2}$. (Here the inhomogeneity is important. If one assumes $T_{C1}<T_{C2}$ the results would be identical.) 

The current through each Josephson junction in the middle of the chain is the sum of quasiparticle and Josephson contributions,
\begin{eqnarray}
I(V) = I_{qp}(V) + I_C\langle \sin\varphi\rangle.
\label{IV}
\end{eqnarray}
Here $V$ is the voltage drop across a single junction, $I_C$ is the critical current of the junction given
by Ambegaokar - Baratoff formula for the junction between two superconductors with different gaps\cite{AB}
\begin{eqnarray}
I_C=\frac{1}{eR_n}\int_{\Delta_1}^{\Delta_2} dE \frac{\Delta_1\Delta_2\tanh\frac{E}{2k_BT}}{\sqrt{E^2-\Delta_1^2}\sqrt{\Delta_2^2-E^2}},
\label{IC}
\end{eqnarray}
and angular brackets denote the averaging over the fluctuations of the phase $\varphi$.
In accordance with Eq. (\ref{IV}), the inverse zero bias resistance of the chain has the form
\begin{eqnarray}
R_0^{-1}(T) = R_{qp}^{-1}(T) + G_J(T),
\label{R_par}
\end{eqnarray}
where
\begin{eqnarray}
\frac{1}{R_{qp}(T)} &=& \frac{1}{N}\frac{\partial I_{qp}(V)}{\partial V}\bigg|_{V=0},
\label{Rqp_def}\\
G_J(T) &=& \frac{1}{N}\frac{\partial}{\partial V}I_C\langle \sin\varphi\rangle\bigg|_{V=0}.
\label{GJ_def}
\end{eqnarray}
Clearly, for temperatures $T>T_{C2}$ the Josephson conductance $G_J$ vanishes and quasiparticle contribution defines the resistance of the chain. 
This is the regime of Giaever tunneling.

\begin{figure*}
\includegraphics[width=1.5\columnwidth]{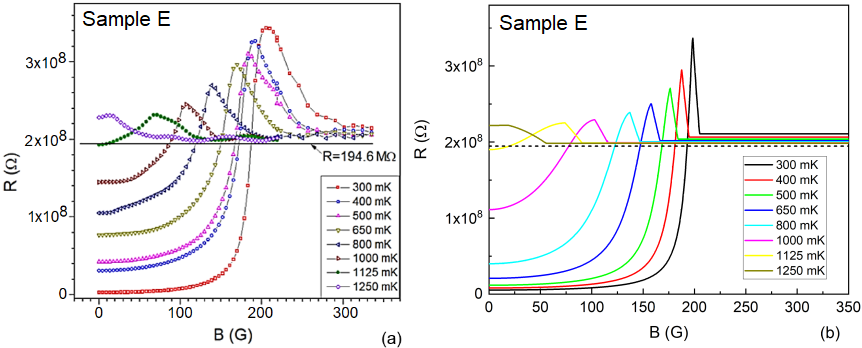} 
\caption{Zero bias resistance of sample E versus magnetic field. Panel a shows the experiment,
panel b -- theory based on Eq. (\ref{R_par},\ref{GJ},\ref{Rqp}).
The parameters used for the theory plots for sample E were
the same as in the caption of Fig. \ref{RT}.}
\label{RH}
\end{figure*}

Since the chain consists of junctions with the resistances of a few k$\Omega$, Coulomb blockade effects should be taken into account.
Let us first consider the quasiparticle contribution to the current $I_{qp}(V)$.
At temperatures $k_BT\gtrsim E_C$, where the effect of random gate potentials induced in the islands by charged impurities
is negligible, one can express the quasiparticle current in the form \cite{IN,Falci}
\begin{eqnarray}
I_{qp}(V) &=& \frac{1}{R_n} \int dE_1dE_2 \, N_1(E_1-eV)N_2(E_2)
\nonumber\\ &&\times\,
\big\{f(E_1-eV)[1-f(E_2)] P_{qp}(E_1-E_2)
\nonumber\\ &&
-\, [1-f(E_1-eV)]f(E_2) P_{qp}(E_2-E_1)\big\}.
\label{Iqp}
\end{eqnarray}
In this expression $f(E)=1/(1+e^{E/k_BT})$ is the Fermi function,
$
N_j(E) = \left|{\rm Re}\left(E/\sqrt{E^2-\Delta_j^2}\right)\right|
$
are the densities of states in the islands of the two types, $R_n$ is the normal state resistance of a single junction, and $P_{qp}(E)$
is the probability of emission of the energy $E$ into the environment during a quasiparticle tunneling event. The function $P_{qp}(E)$
is given by  \cite{IN}
\begin{eqnarray}
P_{qp}(E) = \int\frac{dt}{2\pi\hbar} \, e^{iEt/\hbar}\left\langle e^{i\hat\varphi(t)/2} e^{-i\hat\varphi(0)/2} \right\rangle.
\label{Pqp}
\end{eqnarray}
While evaluating this correlator one should treat the Josephson phase as a quantum operator $\hat\varphi(t)$.
In the limit of highly resistive environment, which is relevant for our experiment, one can make an approximation \cite{IN}
\begin{eqnarray}
P_{qp}(E) = \delta(E-E_C),
\end{eqnarray}
and express the current $I_{qp}(V)$ in terms of "bare" quasiparticle current through the junction, i.e. the current evaluated at
zero charging energy, $E_C=0$,
\begin{eqnarray}
I_{qp}^{(0)}(V) &=& \frac{1}{R_n}\int dE N_1(E-eV)N_2(E)
\nonumber\\ && \times\,
[f(E-eV)-f(E)].
\label{Iqp0}
\end{eqnarray}
Namely, for highly resistive environment one finds
\begin{eqnarray}
I_{qp}(V) = \frac{I_{qp}^{(0)}(V-E_C/e)}{1-e^{(E_C-eV)/k_BT}} + \frac{I_{qp}^{(0)}(V+E_C/e)}{1-e^{(E_C+eV)/k_BT}}.
\label{Iqp1}
\end{eqnarray}
The quasiparticle contribution to zero bias resistance of the chain (\ref{Rqp_def}) then takes the form
\begin{eqnarray}
\frac{1}{R_{qp}(T)} &=& \frac{eI_{qp}^{(0)}(E_C/e)}{2Nk_BT\sinh^2\frac{E_C}{2k_BT}}
\nonumber\\ &&
-\, \frac{2}{N(e^{E_C/k_BT}-1)}\frac{dI_{qp}^{(0)}(E_C/e)}{dV}.
\label{Rqp}
\end{eqnarray}
This expression can be easily evaluated numerically.
It takes three different forms in the temperature intervals $T<T_{C2}$ (superconductor - insulator - superconductor junctions),
$T_{C2}<T<T_{C1}$ (normal metal - insulator - superconductor junctions) and $T>T_{C1}$ (normal metal - insulator - normal metal junctions).
In the latter case and in the limit $k_BT\gg E_C$ Eq. (\ref{Rqp}) reduces to the well known expression
\begin{eqnarray}
R_{qp}(T) = NR_n(1+E_C/3T),
\label{CBT}
\end{eqnarray}
which is used for the Coulomb blockade thermometry \cite{Jukka}.

Next, we consider the average value of the Josephson current.
We first consider the limit $E_J\gg E_C$ at zero temperature.
In this case, in the temperature interval $2E_J/k_B \lesssim T < T_{C2}$ one can use perturbation theory is $E_J$ combined with the theory of
environmental Coulomb blockade \cite{IN}, and express the Josephson current in the form \cite{Averin,Falci}
\begin{eqnarray}
I_C\langle \sin\varphi\rangle = \frac{\pi\hbar I_C^2}{4e}[P_{\rm cp}(2eV)-P_{\rm cp}(-2eV)].
\label{IJ}
\end{eqnarray}
Here
\begin{eqnarray}
P_{\rm cp}(E) = \int\frac{dt}{2\pi\hbar} \, e^{iEt/\hbar}\left\langle e^{i\hat\varphi(t)} e^{-i\hat\varphi(0)} \right\rangle
\label{Pcp}
\end{eqnarray}
is the probability to emit energy $E$ in the environment during a Cooper pair tunneling event in one of the junctions.
The function $P_{\rm cp}(E)$ resembles the function (\ref{Pqp}), but differs from it by a different pre-factor
in front of the phase. This difference arises from the difference between the charge of a single electron ($e$) and that of a Cooper pair ($2e$).
For a junction in highly resistive environment one can approximate \cite{IN}
\begin{eqnarray}
P_{\rm cp}(E) = \frac{\exp\left[-\frac{(E-4E_C)^2}{16E_Ck_BT}\right]}{\sqrt{16\pi E_Ck_BT}}.
\label{Pcp2}
\end{eqnarray}
Taking the derivative of Eq. (\ref{IJ}) at zero bias, we find the Josephson conductance in the form
\begin{eqnarray}
G_J(T) = \frac{\pi\hbar I_C^2}{N}P'_{\rm cp}(0)
= \frac{1}{NR_q} \frac{E_J^2}{E_C^2} \left(\frac{\pi E_C}{k_BT}\right)^{3/2} e^{-\frac{E_C}{k_BT}}.
\nonumber\\
\label{GJ}
\end{eqnarray}
Due to the presence of the factor $e^{-\frac{E_C}{k_BT}}$ this expression exhibits insulating behavior in the limit $T\to 0$.
However, at sufficiently low temperatures, namely for
\begin{eqnarray}
\frac{\sqrt{8E_JE_C}}{2\pi k_B}\lesssim T \lesssim \frac{2E_J}{k_B},
\label{cond1}
\end{eqnarray}
the height of the barrier between the neighboring wells of the cosine Josephson potential, $2E_J$, exceeds the temperature, and
the Josephson conductance becomes determined by thermally activated phase slips. In this case it has characteristic Arrhenius temperature dependence,
\begin{eqnarray}
\frac{1}{G_J(T)} = NR_q \frac{\sqrt{8E_JE_C}}{4k_BT} \, e^{-\frac{2E_J}{k_BT}},
\label{activation}
\end{eqnarray}
which is equivalent to Eq. (\ref{R0}) from the previous section, and should replace Eq. (\ref{GJ}).
The expression (\ref{activation}) implies superconductivity at low temperatures.
In the limit, $E_J\ll E_C$, one should use the expression (\ref{GJ}) for the Josephson conductance down to zero temperature
because the condition (\ref{cond1}) is never satisfied.
If $E_J\sim E_C$, both expressions (\ref{activation}) and (\ref{GJ}) give similar results except for very low temperatures.

Finding the maximum of the average Josephson current
(\ref{IJ}), in which $P_{\rm cp}(E)$ is given by  Eq. (\ref{Pcp2}), one can determine the apparent critical current,
i.e. the maximum current, achieved at $V=2E_C/e$, at which the switching of a junction to the resistive state should occur in the temperature interval $E_C,2E_J<k_BT<k_BT_{C2}$:
\begin{eqnarray}
I_{\rm sw}= \frac{\sqrt{\pi}\hbar I_C^2}{16 e\sqrt{E_Ck_BT}}=\frac{Nk_BT}{2eR_0(T)}.
\label{Isw}
\end{eqnarray}
This expression is applicable when the junction resistance is still close to its normal state value, 
and the phase slips are so frequent that they overlap in time. 
Therefore, this experssion cannot be used at the lowest temperatures of the experiment,
where clear critical current is observed in our V-I curves. Eq. (\ref{Isw}) is presented here for the completeness of the picture.

The presence of magnetic field suppresses superconductivity in each aluminum island due to the Meissner currents. Thus the expressions
for the resistance of the chain given above should be modified.
The full theory of this regime has been developed by Maki \cite{Maki}.
Here we, instead, use a simplified approach, which should be valid
for rather thick films and low magnetic fields used in our experiment.
Namely, we assume that the critical temperature of an island scales with the magnetic field as \cite{Tinkham,Broussard}

\begin{eqnarray}
T_{Cj}(B) = T_{Cj}\left(1-{B^2}/{B_{Cj}^2}\right),
\label{TCH}
\end{eqnarray}
where $B_{Cj}$ are zero-temperature critical fields of the even and odd islands.
We then substitute the critical temperatures (\ref{TCH})
in the approximate gap equation (\ref{DH}), and subsequently evaluate the critical current (\ref{IC}) and the quasiparticle current (\ref{Iqp0})
with the new values of the superconducting gaps $\Delta_1(B)$ and $\Delta_2(B)$.

In Fig. \ref{RT}, we compare theory predictions with the experiment.
The data for the highly resistive sample E are well fitted by Eqs. (\ref{R_par},\ref{Rqp},\ref{GJ}).
For this sample the condition (\ref{cond1}), which is required for the activation approximation for the conductance (\ref{activation})
to be valid, is never truly satisfied, and we used the expression (\ref{GJ}) down to the lowest temperature.
In contrast, for the low resistive sample B we have used both Eq. (\ref{GJ}) and $(\ref{activation})$
for the corresponding temperature intervals. The mismatch between these expressions at $k_BT=2E_J$ produces small jumps in the theoretical curves.
The overall agreement between the theory and the experiment is quite good. All fit parameters for sample B, 
listed in the caption of Fig. \ref{RT}, have been verified independently except for the values of the critical fields $B_{C1}$ and $B_{C2}$. 
For sample E, we also had to reduce the charging energy $E_C$ by a factor of two
as compared to its estimated value.

\begin{figure*}
\includegraphics[width=1.5\columnwidth]{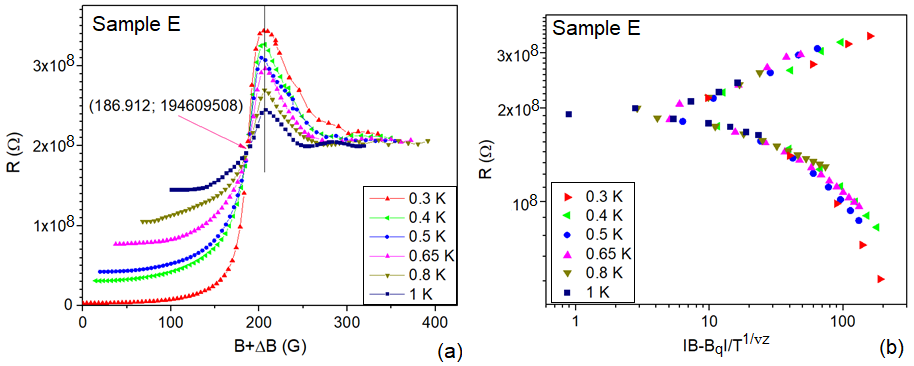}
\caption{ (a) Same data as in Fig.\ref{RH}a, but the curves are shifted along the horizontal axis such that the resistive peak occurs at the same position, namely at the position of the peak of the \textit{T}=300 mK curve. The point where all the measured curves cross indicate a possible critical point, $B_q$ =187 G, $R_C$ = 194.6 MOhm. (b) Scaling of the \textit{R}(\textit{B}) curves, taken at various temperatures, using the scaling relation. The scaling exponent is $\nu z$=0.5. The scaling analysis here does not include points at fields higher than the peak field $B_{c1}$ since at $B>B_{c1}$ vortices are present in the islands. Also, the low-field points were the resistance saturates, presumably due to thermal phase slips, are excluded.}
\label{Fig7}
\end{figure*}

The focus of this paper is the higher-than-normal resistance peak, observed with cooling when the sample just begins developing local superconductivity in the islands. Yet, globally it is not superconducting, and the resistance of the chain increases, as is shown in Fig. \ref{RT}a, Fig. \ref{RT}c. A similar increase of the resistance with cooling has been observed in quench-condensed superconducting films\cite{Dynes}, in granular superconducting films\cite{Adkins}, and in high-$T_C$ superconducting materials which are made of weakly coupled superconducting layers \cite{Briceno,Yurgens}. Thus, this phenomenon is quite general.

In Fig. \ref{RH}, we plot zero-bias resistance of sample E versus magnetic field for different temperatures. The experimental panel (a) also shows a pronounced resistance peak, which gets larger at lower temperatures. The right panel (b) presents the results of the theoretical model outlined above. The agreement between the theory and the experiment is quite good. The peak appears to diverge in the limit of zero temperature due to vanishing tunneling probability in S-I-N junctions. Thus a phase transition in such chains can occur, in principle, even at zero temperature.

\section{Scaling}

In this section we report scaling behavior of the experimental resistance curves, 
which may point to an interesting physics. We do not yet fully understand the origin of this scaling,
although it is clearly related to the transition from a chain of S-I-S' Josephson junctions to a chain of S-I-N Giaever junction
at a certain value of the magnetic field. 

To observe the scaling behavior in the data, we plot a set of \textit{R}(\textit{B}) curves measured at various temperatures (Fig.\ref{RH}a). Then we shift all curves along the horizontal axis so that the peaks are positioned at the same magnetic field for all curves. The shift is denoted $\Delta B$. For the lowest temperature curve (\textit{T}=318 mK; red squares) $\Delta B$=0. As such shifts are finalized, the critical point $B_q$ emerges on such plot as a crossing point for all curves, see Fig.\ref{Fig7}a. The crossing of all the measured curves at the same point suggests that this point can be considered a critical point.

Let us first discuss the peak, and then it will be easier to understand the crossing point. The resistance maximum (the peak in Fig.\ref{Fig7}a) occurs at the field which suppresses the BCS condensate in the islands with lower $T_C$, which we have denoted as $B_{C2}$. A higher field, at which the resistance approaches the normal resistance $R_n$, is the critical field of the islands with higher $T_C$, namely, $B_{C1}$.

If we assume that the crossing point is the critical point of a phase transition at zero temperature, then we expects that, as the temperature approaches zero, the slope of the curve approaches infinity, namely d\textit{R}/d\textit{B}$\rightarrow$$\infty$ at  $B=B_q$. On the other hand, the resistance peak height increases with cooling (which will be shown explicitly in the next section), so its maximum, positioned at \textit{B}=$B_{C2}$, should also approach infinite resistance in the limit of zero temperature. Interestingly, the peak appears at a fixed separation, about 25 G, from the crossing point (Fig.\ref{Fig7}a). Therefore, one concludes that, in the limit of \textit{T}$\rightarrow$0, the resistance in the range $B_q<B<B_{C2}$ should be infinity, \textit{R}(\textit{B})$\rightarrow$$\infty$. Thus the most natural way to understand the critical (crossing) point in the family of $R(B)$ curves as the point where the even-numbered islands experience a transition between normal and superconducting states. At fields and temperatures below this point, the condensate begins to flow along the chain, on the time scale defined by the time intervals between consecutive phase slips.

Phase transitions are known to exhibit scaling behavior.
The scaling analysis of Fig.\ref{Fig7}b is based on the well-known resistance scaling equation\cite{Sondhi,Fisher,Hebard}
\begin{eqnarray}
R_0(T,B)=F\left(\frac{|B-B_q|}{T^{\nu z}}\right).
\label{scaling} 
\end{eqnarray}
Here $F$ is an unknown function, which has two branches, one representing the superconducting regime and the other corresponding to the insulating regime. The product of the space and time correlation exponents, $\nu z$, is chosen to produce the best possible collapse of the experimental curves on the universal scaling function F. In our case, the best value is $\nu z$=0.5. The collapse of the curves provides evidence that a phase transition occurs in the chain when even number islands 
become normal. Note that it was recently demonstrated that the exact type of the phase transition cannot be determined from the scaling analysis along because models developed for two dimensional systems appear to be applicable to one dimensional samples such as superconducting nanowires\cite{Rogachev}. 

Theoretical model of Sec. \ref{theory} 
does not show the exact scaling property Eq. (\ref{scaling}), it agrees with Eq. (\ref{scaling}) only roughly. 
The model states, for example, that zero bias resistance of the array
should be a function of three dimensionless parameters: $\Delta_1(B)/k_BT$, $\Delta_2(B)/k_BT$ and $E_C/k_BT$.
The ratios $\Delta_1(B)/k_BT$ and $\Delta_2(B)/k_BT$ can be further expressed in terms of the two  dimensionless combinations
\begin{eqnarray}
x_{j} = \frac{T_{Cj}(B)}{T} \equiv \frac{T_{Cj}}{T}-\frac{T_{Cj}}{B_{Cj}^2}\frac{B^2}{T},\;\; j=1,2.
\end{eqnarray}
Here we have used the experssion (\ref{TCH}) for the critical temperatures of the islands in presence of the field.
Thus, magnetic field always appears in the theory as the ratio $B/\sqrt{T}$, which is in agreement with the
value of the scaling parameter $\nu z=0.5$ discussed above. Two scaling parameters can emerge if the magnetic field is near the critical field, corresponding to zero temperature, of one of the islands. Then the gap in the even islands would be approaching zero while the gap in the odd islands would remain strong and could be treated as a constant near such field-driven transition.

\section{Extrapolation to zero temperature}

In what follows, we will argue, phenomenologically, that below a specific field, the resistance peak can grow so large that it will cover a temperature interval extended down to zero temperature. The magnetic field at which the resistance peak maximum coincides with the point \textit{T}=0 could present a critical point of a quantum superconductor-insulator transition (SIT) for a chain of Giaever junctions (i.e., S-I-N junctions).
For the evidence of such behavior one can look at the $R(T)$-curves measured at 156 G for sample B and at 250 G for sample E (Fig.\ref{RT}a and Fig.\ref{RT}c). These curves exhibit growing resistance in the entire explored temperature interval. Note that if the magnetic field is made much higher than these values, then the local superconductivity within each grain becomes suppressed and the resistance peak disappears. This fact confirms that the resistance peak is related to the superconductivity in the islands. 

\begin{figure}
\includegraphics[width=\columnwidth]{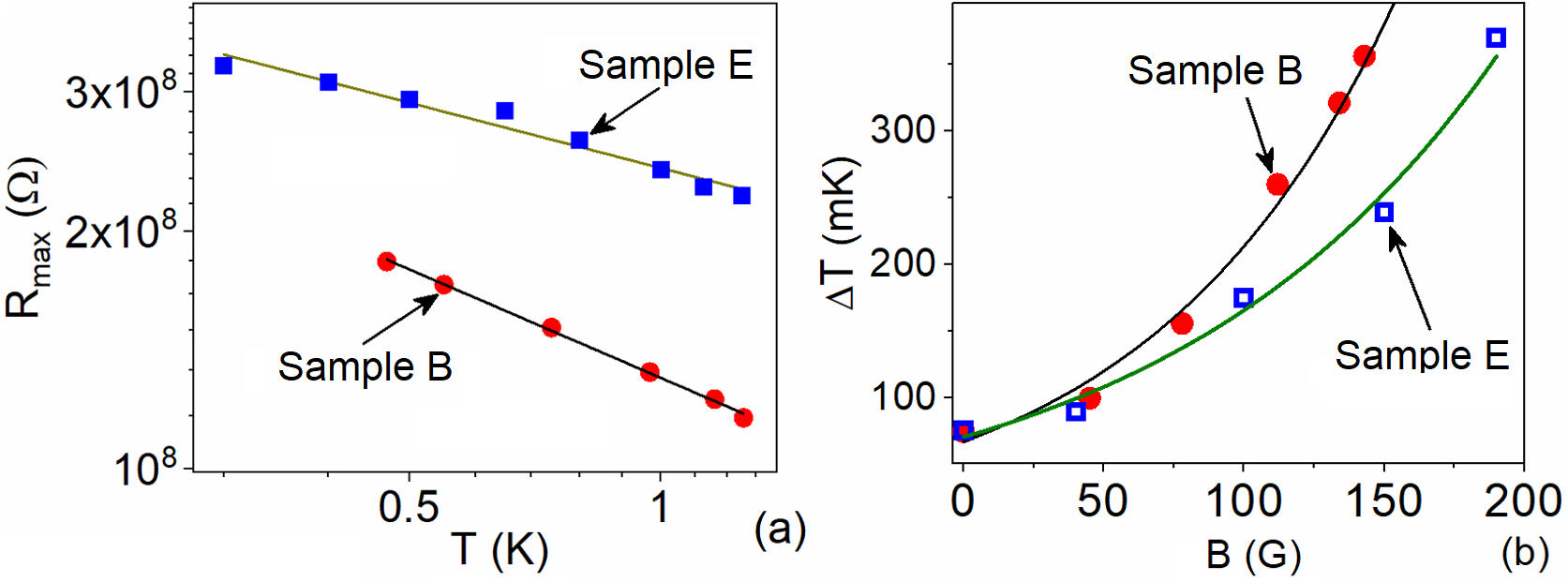}
\caption{(a) The peak resistance value, $R_{\max}$, is plotted versus the temperature at which the peak occurs, during a cool-down with the magnetic field being fixed. The linear fits are given by the formulas $R_{\max}=0.24GOhm/(T/1K)^{0.27}$ and $R_{\max}=0.13GOhm/(T/1K)^{0.45}$ for the samples E and B correspondingly. Both the vertical and the horizontal scales are logarithmic. The $R_{\max}$ positions are shown by arrows in Fig.\ref{RT}a and Fig.\ref{RT}c. (b) Temperature dependence of the peak half-width, $\Delta T=T_{in}-T_{\max}$, is plotted versus the magnetic field. It was measured from the higher-temperature side of the resistance peak to the resistance maximum. 
The formula for the fit is $\Delta T=\Delta T(0)(1+B^2/B_1^2)$, where $\Delta T$(0)=67.68 mK and $B_1$=68.58 G for sample B, 
and $\Delta T$(0)=71.52 mK, $B_1$=98.32 G for sample E.}
\label{Fig5}
\end{figure}

\begin{figure}
\includegraphics[width=\columnwidth]{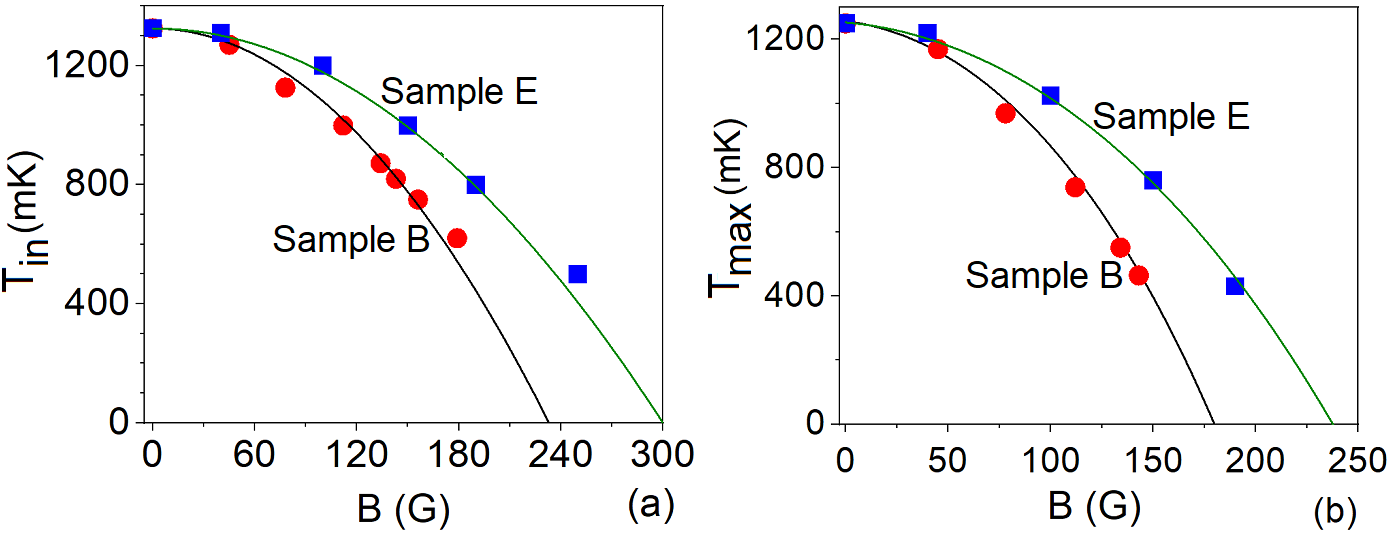}
\caption{(a) The beginning of the resistance peak, $T_{in}$, plotted versus magnetic field, for the two samples. The formula for the fit is $T_{in}(B)=T_{in}(0)(1-B^2/\tilde B_{in}^2)$, where $T_{in}$(0)=1325 mK, $\tilde B_{in}$=233 G for sample B (red circles), and $T_{in}$(0)= 1325 mK, 
$\tilde B_{in}$ =300 G for sample E (blue squares). (b) The position of the maximum of the  resistance peak, $T_{\max}$, plotted as a function 
of the magnetic field. The formula for the fit is $T_{\max}(B)=T_{\max}(0)(1-B^2/\tilde B_{\max}^2)$. The fit parameters are
$T_{\max}(0)=1257$ mK, $\tilde B_{\max}=180$ G for the sample B (red circles); 
and $T_{\max}(0)=1253$ mK, $\tilde B_{\max}=238$ G for the sample E (blue squares).}
\label{Fig6}
\end{figure}

We now analyze the resistance peak and show that, as the magnetic field is increased, and the system approaches zero-temperature field-controlled transition, the peak exhibits and accelerated growth, both in height and in width. The normal resistance $R_n$ is defined as the sample resistance at high temperature ($T>T_{in}$), and $R_{\max}$ is the resistance at the maximum of the peak. Here $T_{in}$ is the temperature at which the resistance begins to increase with cooling, see Fig.\ref{RT}a and Fig.\ref{RT}c. 
Is is clear from the log-log plots in Fig.\ref{Fig5}a that the peak height, $R_{max}$, exhibits a power-law growth with the temperature at which the peak occurs. This is due to the electron transfer suppression between the normal and the superconducting grains in S-I-N junctions at low temperatures. Our chains can mimic disordered superconductors in which the local value of the critical field fluctuates considerably, and the observed saturation of the resistance indicates an insulating regime in such S-I-N structures. The theory model of Sec. \ref{theory} leads to a similar dependence of the peak height on temperature, as it is evident from Figs. \ref{RT}b,d. 
Unfortunately, no simple analytical expression can be derived for the range of parameters of Fig. \ref{Fig5}a. The model predicts, however, a divergence
of the peak height as the magnetic field approaches the critical field of the islands with weaker superconductivity.
Moreover, the peak position in the limit $B\to B_{C2}$ shifts towards zero temperature.
Thus, the superconductor - insulator phase transition may indeed occur at the magnetic field $B_{C2}$, at which the superconductivity in the weaker islands is
suppressed.

We also analyze the width of the resistance 
peak as a function of the applied magnetic field. We define it as the half-width of the peak, $\Delta T$=$T_{in}$-$T_{\max}$, where 
$T_{in}$ is the temperature at which the resistance begins to grow (in a cool-down), and $T_{\max}$ is the temperature at which the resistance reaches its maximum. 
The result of this analysis is shown in Fig. \ref{Fig6}. 
The temperatures $T_{in}$ and $T_{\max}$ have parabolic dependence on the magnetic field, $T_{in}(B)=T_{in}(0)(1-B^2/\tilde B_{in}^2)$ 
and $T_{\max}(B)=T_{\max}(0)(1-B^2/\tilde 
B_{\max}^2)$.
According to the model, we expect $T_{in}=T_{C1}(B)$ and $T_{\max}=T_{C2}(B)$ and the critical temperatures should, indeed, 
have parabolic dependence on magnetic field (\ref{TCH}). The fitting gives the values of the fields $\tilde B_{in}$ and $\tilde B_{\max}$
similar, but not equal to the critical fields $B_{C1}$ and $B_{C2}$ used for the theory plots in Fig. \ref{RT}b,d. 
This discrepancy may be explained by broadening of the experimental resistance peaks, which complicates
precise determination of the values of the fields $B_{C1}$ and $B_{C2}$.  
Finally, the peak width $\Delta T=T_{in}-T_{\max}$, plotted in Fig. \ref{Fig5}b,  also has quadratic dependence on the magnetic field.

In the end of this section, 
we would like to note that the superconductor - insulator phase transition, which we discussed above, differs from the commonly discussed one \cite{BD,Lee}. 
The latter phase transition is caused by partial suppression of the Josephson energy of a single junction, and at
the critical field  $E_J$ becomes of the order of $E_{Cg}$. At this field all islands in the chain still remain superconducting. 
The interplay between the two phase transitions may be complicated. Indeed, as we have discussed above, our arrays should formally be insulating at zero temperature.
However, this insulating behavior can only be observed at very low, inaccessible, temperatures.
In contrast, the transition to the chain of S-I-N junctions at $B=B_{C2}$ can be visible already at rather high temperatures.
On the other hand, a chain of S-I-N junctions is never truly insulating in the absence
of Coulomb blockade because Andreev tunneling and elastic cotunneling through the normal islands
limit growth of the resistance at low temperatures.
We hope our results will initiate further in-depth theoretical analysis of S-I-N chains involving these issues.

\section{Conclusion}

In conclusion, we study phase transitions in chains of weakly coupled superconducting islands that have alternating critical temperatures and alternating critical fields. If the temperature and/or magnetic field is such that superconductivity is suppressed in all islands, then all islands are normal, and the chain behaves like a normal metal. If the temperature and/or magnetic field is such that superconductivity is suppressed in all even islands, while all odd islands remain internally superconducting, then the chain behaves as a sequence of Giaever junctions (i.e., S-I-N junctions), and is insulating in the limit of zero temperatures (if Andreev reflection and co-tunneling are negligible). The experimental manifestation of this insulating behavior is the observed resistance peak. At the peak, the chain resistance goes higher than its resistance in the normal state. Finally, if the temperature and magnetic field are low enough, all islands become superconducting, and the chain acts as a sequence of S-I-S junctions, which is superconducting, with resistance dropping roughly exponentially with cooling.

Our results are directly applicable to systems and materials in which small superconducting grains are separated by strong barriers, such as oxidized layers for example. We speculate that under certain conditions an analogous behavior might even occur in homogeneous amorphous metallic films if, due to strong disorder, a significant fraction of the film becomes gapped. In this respect, it is interesting to consider the pioneering STM (scanning tunneling microscopy) spectroscopy studies of the superconducting gap in strongly disordered thin films, performed by the Sacépé and collaborators \cite{SacepeDisorder, SacepeSI}. They demonstrated that superconductivity is not completely suppressed at the critical disorder but persists into the insulating regime in the form of localized superconducting “lakes” or “islands”. The key fact linking the STM study results and our present results, is that insulating regions and islands have been also observed, even the homogeneous but strongly disordered films. Thus, an interesting possibility exists that those insulating regions might provide sufficiently strong barriers between the superconducting regions for the creation of S-I-N patterns. Additional research is needed to make a definite conclusion about such possibility. In case if the strongly disordered films can indeed spontaneously develop insulating regions which play the role of the barriers, then the temperature and/or the magnetic field could be adjusted such that superconducting, normal, and insulating regions would form S-I-N junctions even at zero temperature. Then, due to the Giaever resistance divergence, as insulating regime can emerge. We hope our model system will initiate further research into this type of mechanism, generating insulating or highly resistive regimes in granular superconducting systems, in which superconducting grains are separated by sufficiently strong insulating barriers. Its applicability to strongly disordered but homogeneous films remains hypothetical.

The conductivity of an S-I-N- infinite chain in the limit of zero temperature remains an open theoretical problem. Although Andreev reflection and co-tunneling could generate some conductivity, so the chain would not be fully insulating, the Coulomb charging phenomena could render the chain insulating. Further transport measurements on such arrays and systems with the critical temperature fluctuations, at lower temperatures, are also warranted.

\section{Acknowledgments}
We would like to acknowledge illuminating conversations with V. Manucharyan and J. Aumentado. The samples were kindly provided by V. Manucharyan group. This work was supported by NSF DMR 18-36710.
This project has also received funding from the European Union's Horizon 2020 research and innovation program under grant agreement No 862660/QUANTUM E-LEAPS.

\end{document}